\documentstyle[12pt]{article}

\newif\ifprelim

\def\mode{b }

\def\baselinestretch{1.2}
\def\unlock{\catcode`@=11 }
\def\lock{\catcode`@=12 }

\def\marginnote#1{}
\let\ju=\marginnote
\def\mybiblabel#1{#1\hfil}
\unlock
\def\@bibitem#1{\def\@blbl{#1}\item\if@filesw \immediate\write\@auxout
        {\string\bibcite{#1}{\the\value{\@listctr}}}\fi\ignorespaces}
\lock
\def\blackfonts{
                \font\blackboard=msbm10 scaled\magstep1
                \font\blackboards=msbm8
                \font\blackboardss=msbm6}

\def\bigmode{b }
\def\draftmode{d }
\def\Draftmode{D }

\def\Newif#1{\expandafter\ifx\csname#1\endcsname\relax
 \csname newif\expandafter\endcsname\csname#1\endcsname\fi}
\Newif{ifprelim}

\ifx\mode\undefined\read-1 to\mode\fi
\def\bigpage{
        \textheight 22.5 cm
        \topmargin -.5 cm
        \textwidth 16cm
        \oddsidemargin 0 in
        \evensidemargin 0 in
}
\def\doublepage{
        \parskip 6 pt
        \parindent 2em
        \twocolumn
        
        \let\small\relax
        \let\sl\it
        \sloppy
        \voffset=-2.54truecm
        \hoffset=-1.54truecm
        \flushbottom
        \leftmargini 2em
        \leftmarginv .5em
        \leftmarginvi .5em
        \marginparwidth 48pt
        \marginparsep 10pt
        \setlength{\columnsep}{2truecm}
        \setlength{\textwidth}{25.4truecm}
        \setlength{\textheight}{17truecm}
        \oddsidemargin .18truein
        \evensidemargin .17truein
}
\def\draftpage{
        \textheight 19.86075cm
        \topmargin -.5 cm
        \textwidth 13.66667cm
        \oddsidemargin 0cm
        \evensidemargin 0cm
        \marginparwidth 96pt
        \marginparsep 10pt
}
\def\smallblack{
        \def\blackfonts{
                \font\blackboard=msbm10
                \font\blackboards=msbm7
                \font\blackboardss=msbm5
        }
}
\def\marginnotes{
        \def\draftmarginnote##1{\marginpar{\raggedright\scriptsize\tt ##1}}
        \let\marginnote=\draftmarginnote
        \let\ju=\marginnote
}
\unlock
\def\draftnotice{
        \def\@oddhead{\hfil \smash{\Large\em DRAFT} \ \ --- \
                \em \today\quad\militarytime 
        \hfil}
        \let\@evenhead\@oddhead
        \def\ps@plain{\let\@mkboth\@gobbletwo
                \def\@oddfoot{\hfil 
                \thepage 
                \hfil}
                \let\@evenfoot\@oddfoot}
        \def\ps@empty{\let\@mkboth\@gobbletwo
                \def\@oddfoot{\hfil \smash{\Large\em DRAFT} \hfil}
                \let\@evenfoot\@oddfoot}
        \pagestyle{plain}
}
\def\eqnlabels{
        \def\draftlabel##1{{\@bsphack\if@filesw {\let\thepage\relax
           \xdef\@gtempa{\write\@auxout{\string
              \newlabel{##1}{{\@currentlabel}{\thepage}}}}}\@gtempa
           \if@nobreak \ifvmode\nobreak\fi\fi\fi\@esphack}
                \gdef\@eqnlabel{##1}}
        \def\@eqnlabel{}
        \def\@vacuum{}
        \let\label=\draftlabel
        \def\@eqnnum{(\theequation)\rlap{\kern\marginparsep\tt\@eqnlabel}%
                \global\let\@eqnlabel\@vacuum}
}
\def\draftbib{
        \def\mybiblabel##1{\llap{\scriptsize\tt \@blbl\ }##1\hfil}
        \def\@lbibitem[##1]##2{%
                \def\@blbl{##2}\item[\@biblabel{##1}\hfill]\if@filesw
                {\def\protect##1{\string ##1\space}\immediate
                \write\@auxout{\string\bibcite{##2}{##1}}}\fi\ignorespaces}
        \def\@bibitem##1{\def\@blbl{##1}\item\if@filesw
                \immediate\write\@auxout
                {\string\bibcite{##1}{\the\value{\@listctr}}}\fi\ignorespaces}
        }
\lock

\ifx\mode\bigmode
        \ifprelim
                \draftnotice
        \else
        \fi
        \bigpage
\else\ifx\mode\draftmode
        \draftpage
        \draftnotice
        \marginnotes
        \eqnlabels
        \draftbib
\else\ifx\mode\Draftmode
        \bigpage
        \draftnotice
        \marginnotes
        \eqnlabels
        \draftbib
\else                           
        \ifprelim
                \draftnotice
        \else
        \fi
        \unlock
        \input art10.sty
        \lock
        \def\baselinestretch{1.3}
        \smallblack
        \doublepage
\fi
\fi
\fi

\batchmode
        \newfont{\footbbbfont}{msbm10}
\errorstopmode

\newif\ifamsf\amsftrue
        \ifx\footbbbfont\nullfont
        \amsffalse
\fi

\ifamsf
        \blackfonts
        \newfam\black
        \textfont\black=\blackboard
        \scriptfont\black=\blackboards
        \scriptscriptfont\black=\blackboardss

\else            

\fi

\newcount\hour
\newcount\minute
\newtoks\amorpm
\hour=\time\divide\hour by 60
\minute=\time{\multiply\hour by 60 \global\advance\minute by-\hour}
\edef\standardtime{{\ifnum\hour<12 \global\amorpm={am}%
        \else\global\amorpm={pm}\advance\hour by-12 \fi
        \ifnum\hour=0 \hour=12 \fi
        \number\hour:\ifnum\minute<10 0\fi\number\minute\the\amorpm}}
\edef\militarytime{\number\hour:\ifnum\minute<10 0\fi\number\minute}

\newif\ifepsfloaded
\newif\iffigureexists

\ifeof 1 \epsfloadedfalse \else \epsfloadedtrue \fi
\ifepsfloaded \input epsf \fi

\def\checkex#1 {\relax
    \openin 1 #1
    \ifeof 1 \figureexistsfalse
    \else \figureexiststrue
    \fi \closein 1 }

\def\cpsbox#1#2{
        \ifepsfloaded
                \checkex #2
                \iffigureexists
                        \immediate\write16{(#2)}
                        \setlength{\epsfxsize}{#1}
                        \centerline{\epsfbox{#2}}
                \else
                        \immediate\write16{(#2 NOT FOUND!)}
                        \vbox to 2in{\hbox to #1 {\hss} \vss}
                \fi
        \else
                \immediate\write16{(NOT inputting #2; no epsf.tex)}
                \vbox to 2in{\hbox to #1 {\hss} \vss}
        \fi}
\unlock
\def\@citex[#1]#2{%
\if@filesw \immediate \write \@auxout {\string \citation {#2}}\fi
\@tempcntb\m@ne \let\@h@ld\relax \def\@citea{}%
\@cite{%
  \@for \@citeb:=#2\do {%
    \@ifundefined {b@\@citeb}%
      {\@h@ld\@citea\@tempcntb\m@ne{\bf ?}%
      \@warning {Citation `\@citeb ' on page \thepage \space undefined}}%
      {\@tempcnta\@tempcntb \advance\@tempcnta\@ne%
      \@tempcntb\number\csname b@\@citeb \endcsname \relax%
      \ifnum\@tempcnta=\@tempcntb 
        \ifx\@h@ld\relax%
          \edef \@h@ld{\@citea\csname b@\@citeb\endcsname}%
        \else%
          \edef\@h@ld{\ifmmode{-}\else--\fi\csname b@\@citeb\endcsname}%
        \fi%
      \else
        \@h@ld\@citea\csname b@\@citeb \endcsname%
        \let\@h@ld\relax%
      \fi}%
    \def\@citea{,\penalty\@highpenalty\,}%
  }\@h@ld
}{#1}}

\def\@citeb#1#2{{[#1]\if@tempswa , #2\fi}}
\def\@citeu#1#2{{$^{#1}$\if@tempswa , #2\fi }}
\def\@citep#1#2{{#1\if@tempswa , #2\fi}}

\def\bcites{         
        \unlock
        \let\@cite=\@citeb
        \lock
}

\def\upcites{         
        \unlock
        \let\@cite=\@citeu
        \lock
}

\def\plaincites{      
        \unlock
        \let\@cite=\@citep
        \lock
}

\let\@cite=\@citeb              

\def\refname{References}        

\def\thebibliography#1{\section*{\refname\@mkboth
  {\uppercase{\refname}}{\uppercase{\refname}}}\list
  {\@biblabel{\arabic{enumiv}}}{\settowidth\labelwidth{\@biblabel{#1}}%
    \let\makelabel\mybiblabel\leftmargin\labelwidth
    \advance\leftmargin\labelsep
    \usecounter{enumiv}%
    \let\p@enumiv\@empty
    \def\theenumiv{\arabic{enumiv}}}%
    \def\newblock{\hskip .11em plus.33em minus.07em}%
    \sloppy\clubpenalty4000\widowpenalty4000
    \sfcode`\.=1000\relax}

\def\@noitemerr{\@warning{Something's wrong--perhaps a missing
\string\item}\@ehc}

\def\sections{\unlock
\def\theequation{\thesection.\arabic{equation}}
\@addtoreset{equation}{section}
\@addtoreset{footnote}{section}
\lock
}

\def\footnotesections{\unlock
\@addtoreset{footnote}{section}
\lock
}

\def\subsections{\unlock
\def\theequation{\thesubsection.\arabic{equation}}
\@addtoreset{equation}{subsection}
\@addtoreset{footnote}{subsection}
\lock
}

\lock

\def\ibid{{\it ibid.\/}}

\def\Kahler{K\"ahler}
\def\kahler{k\"ahler}
\def\einbein{{\it einbein\/}}

\def\zweibein{{\it zweibein\/}}

\def\CY{Calabi--Yau}
\def\CS{Chern--Simons}

\def\noj#1,#2,{{\bf #1} (19#2)\ }
\def\jou#1#2,#3,{{\em #1\/ }{\bf #2} (19#3)\ }
\def\ann#1,#2,{{\em Ann.\ Physics\/ }{\bf #1} (19#2)\ }
\def\annmath#1,#2,{{\em Ann.\ Math\/ }{\bf #1} (19#2)\ }
\def\cmp#1,#2,{{\em Comm.\ Math.\ Phys.\/ }{\bf #1} (19#2)\ }
\def\cq#1,#2,{{\em Class.\ Quantum Grav.\/ }{\bf #1} (19#2)\ }
\def\cqg#1,#2,{{\em Class.\ Quantum Grav.\/ }{\bf #1} (19#2)\ }
\def\ijmp#1,#2,{{\em Int.\ J.\ Mod.\ Phys.\/ }{\bf A#1} (19#2)\ }
\def\jmp#1,#2,{{\em J.\ Math.\ Phys.\/ }{\bf #1} (19#2)\ }
\def\grg#1,#2,{{\em Gen.\ Rel.\ Grav.\/ }{\bf #1} (19#2)\ }
\def\mpl#1,#2,{{\em Mod.\ Phys.\ Lett.\/ }{\bf A#1} (19#2)\ }
\def\nc#1,#2,{{\em Nuovo Cim.\/ }{\bf #1} (19#2)\ }
\def\np#1,#2,{{\em Nucl.\ Phys.\/ }{\bf B#1} (19#2)\ }
\def\pl#1,#2,{{\em Phys.\ Lett.\/ }{\bf #1B} (19#2)\ }
\def\pla#1,#2,{{\em Phys.\ Lett.\/ }{\bf #1A} (19#2)\ }
\def\pr#1,#2,{{\em Phys.\ Rev.\/ }{\bf #1} (19#2)\ }
\def\prd#1,#2,{{\em Phys.\ Rev.\/ }{\bf D#1} (19#2)\ }
\def\prl#1,#2,{{\em Phys.\ Rev.\ Lett.\/ }{\bf #1} (19#2)\ }
\def\prp#1,#2,{{\em Phys.\ Rept.\/ }{\bf #1C} (19#2)\ }
\def\ptp#1,#2,{{\em Prog.\ Theor.\ Phys.\/ }{\bf #1} (19#2)\ }
\def\ptpsup#1,#2,{{\em Prog.\ Theor.\ Phys.\/ Suppl.\/ }{\bf #1} (19#2)\ }
\def\rmp#1,#2,{{\em Rev.\ Mod.\ Phys.\/ }{\bf #1} (19#2)\ }
\def\yadfiz#1,#2,#3[#4,#5]{{\em Yad.\ Fiz.\/ }{\bf #1} (19#2) #3
[{\em Sov.\ J.\ Nucl.\ Phys.\/ }{\bf #4} (19#2) #5]}
\def\zh#1,#2,#3[#4,#5]{{\em Pis'ma\ Zh.\ Exp.\ Theor.\ Fiz.\/ }{\bf #1}
(19#2) #3 [{\em Sov.\ Phys.\ JETP\/ }{\bf #4} (19#2) #5]}

\def\eq#1{.~(\ref{#1})}
\def\noeq#1{(\ref{#1})}
\def\beq{\begin{equation}}
\def\eeq{\end{equation}}
\def\beqar{\begin{eqnarray}}
\def\eeqar{\end{eqnarray}}

\def\beqal{\begin{equation}\begin{eqalign}}
\def\eeqal{\end{eqalign}\end{equation}}
\def\beqaltwo{\begin{eqaligntwo*}}
\def\eeqaltwo{\end{eqaligntwo*}}

\def\p#1{\mskip#1mu}
\def\n#1{\mskip-#1mu}
\def\stop{\p6.}
\def\comma{\p6,}
\def\semi{\p6;}

\def\nfrac#1#2{{\displaystyle{\vphantom1\smash{\lower.5ex\hbox{\small$#1$}}%
        \over\vphantom1\smash{\raise.25ex\hbox{\small$#2$}}}}}
\def\ket#1{\left| #1 \right\rangle}

\def\pa{\partial}

\def\Tr{{\rm Tr}}
\def\l:{\mathopen{:}\,}
\def\r:{\,\mathclose{:}}
\def\diag{{\rm diag}}
\def\det{\mathop{\rm det}\nolimits}

\unlock
\def\@versim#1#2{\smash{\lower0.5ex\vbox{\baselineskip\z@skip\lineskip\z@skip
        \lineskiplimit\z@\ialign{$\m@th#1\hfil##\hfil$\crcr#2\crcr\sim\crcr}}}}
\def\ltsim{\mathrel{\mathpalette\@versim<}}
\def\gtsim{\mathrel{\mathpalette\@versim>}}
\lock

\def\warboxp#1{\setbox0=\hbox{$#1M$}\mkern1.5mu
        \vbox{\hrule height0pt depth.04\ht0
        \hbox{\vrule width.04\ht0 height.9\ht0 \kern.9\ht0
        \vrule width.04\ht0}\hrule height.04\ht0}\mkern1.5mu}
\def\warbox{{\mathpalette\warboxp{}}}                        

\def\boxes#1{
        \newcount\num
        \num=1
        \newdimen\downsy
        \downsy=-1.5ex
        \mkern-3.5mu
        \warbox
        \loop
        \ifnum\num<#1
        \llap{\raise\num\downsy\hbox{$\warbox$}}
        \advance\num by1
        \repeat}
\def\boxup#1#2{\newcount\numup
        \numup=#1
        \newdimen\upsy
        \upsy=.75ex
        \mkern3.5mu
        \raise -.25ex\hbox{{\raise\numup\upsy\hbox{$#2$}}}}

\def\cald{{\cal D}}

\def\calf{{\cal F}}

\def\call{{\cal L}}
\def\calm{{\cal M}}

\def\mubar{{\bar \mu}}
\def\nubar{{\bar \nu}}
\def\xibar{{\bar \xi}}

\def\rhobar{{\bar \rho}}
\def\sigmabar{{\bar \sigma}}

\unlock
\def\section{\@startsection {section}{1}{\z@}{3.ex plus 1ex minus
 .2ex}{2.ex plus .2ex}{\large\bf}}
\def\subsection{\@startsection{subsection}{2}{\z@}{2.75ex plus 1ex minus
 .2ex}{1.5ex plus .2ex}{\bf}}
\if@twocolumn
                
                \def\abstractindent{1.5 cm}
\else
                
                \def\abstractindent{.75 cm}
\fi

\def\appendix{{\newpage\section*{Appendices}}\let\appendix\section%
        {\setcounter{section}{0}
        \gdef\thesection{\Alph{section}}}\section}

\def\abstract{
        \vfill\begin{center}
        {\bf Abstract}
        \end{center}
        \advance\leftskip\abstractindent
        \advance\rightskip\abstractindent
}

\lock

\unlock

\newif\if@defeqnsw \@defeqnswtrue

\def\eqnarray{\stepcounter{equation}\let\@currentlabel=\theequation
\if@defeqnsw\global\@eqnswtrue\else\global\@eqnswfalse\fi
\global\@eqnswtrue
\tabskip\@centering\let\\=\@eqncr
$$\halign to \displaywidth\bgroup\hfil\global\@eqcnt\z@
  $\displaystyle\tabskip\z@{##}$&\global\@eqcnt\@ne
  \hfil$\displaystyle{{}##{}}$\hfil
  &\global\@eqcnt\tw@ $\displaystyle{##}$\hfil
  \tabskip\@centering&\llap{##}\tabskip\z@\cr}

\def\yesnumber{\global\@eqnswtrue}

\def\@@eqncr{\let\@tempa\relax\global\advance\@eqcnt by \@ne
    \ifcase\@eqcnt \def\@tempa{& & & &}\or \def\@tempa{& & &}\or
     \def\@tempa{& &}\or \def\@tempa{&}\else\fi
     \@tempa \if@eqnsw\@eqnnum\stepcounter{equation}\fi
     \if@defeqnsw\global\@eqnswtrue\else\global\@eqnswfalse\fi
     \global\@eqcnt\z@\cr}

\def\@eqnacr{{\ifnum0=`}\fi\@ifstar{\@yeqnacr}{\@yeqnacr}}

\def\@yeqnacr{\@ifnextchar [{\@xeqnacr}{\@xeqnacr[\z@]}}

\def\@xeqnacr[#1]{\ifnum0=`{\fi}\cr \noalign{\vskip\jot\vskip #1\relax}}

\def\eqalign{\null\,\vcenter\bgroup\openup1\jot \m@th \let\\=\@eqnacr
\ialign\bgroup\strut
\hfil$\displaystyle{##}$&$\displaystyle{{}##}$\hfil\crcr}
\def\endeqalign{\crcr\egroup\egroup\,}

\def\cases{\left\{\,\vcenter\bgroup\normalbaselines\m@th \let\\=\@eqnacr
    \ialign\bgroup$##\hfil$&\quad##\hfil\crcr}
\def\endcases{\crcr\egroup\egroup\right.}

\def\eqalignno{\stepcounter{equation}\let\@currentlabel=\theequation
\if@defeqnsw\global\@eqnswtrue\else\global\@eqnswfalse\fi
\let\\=\@eqncr
$$\displ@y \tabskip\@centering \halign to \displaywidth\bgroup
  \global\@eqcnt\@ne\hfil
  $\@lign\displaystyle{##}$\tabskip\z@skip&\global\@eqcnt\tw@
  $\@lign\displaystyle{{}##}$\hfil\tabskip\@centering&
  \llap{\@lign##}\tabskip\z@skip\crcr}

\def\endeqalignno{\@@eqncr\egroup
      \global\advance\c@equation\m@ne$$\global\@ignoretrue}

\@namedef{eqalignno*}{\@defeqnswfalse\eqalignno}
\@namedef{endeqalignno*}{\endeqalignno}

\def\eqaligntwo{\stepcounter{equation}\let\@currentlabel=\theequation
\if@defeqnsw\global\@eqnswtrue\else\global\@eqnswfalse\fi
\let\\=\@eqncr
$$\displ@y \tabskip\@centering \halign to \displaywidth\bgroup
  \global\@eqcnt\m@ne\hfil
  $\@lign\displaystyle{##}$\tabskip\z@skip&\global\@eqcnt\z@
  $\@lign\displaystyle{{}##}$\hfil\qquad&\global\@eqcnt\@ne
  \hfil$\@lign\displaystyle{##}$&\global\@eqcnt\tw@
  $\@lign\displaystyle{{}##}$\hfil\tabskip\@centering&
  \llap{\@lign##}\tabskip\z@skip\crcr}

\def\endeqaligntwo{\@@eqncr\egroup
      \global\advance\c@equation\m@ne$$\global\@ignoretrue}

\@namedef{eqaligntwo*}{\@defeqnswfalse\eqaligntwo}
\@namedef{endeqaligntwo*}{\endeqaligntwo}

\newtoks\@stequation

\def\subequations{\refstepcounter{equation}%
  \edef\@savedequation{\the\c@equation}%
  \@stequation=\expandafter{\theequation}
  \edef\@savedtheequation{\the\@stequation}
  \edef\oldtheequation{\theequation}%
  \setcounter{equation}{0}%
  \def\theequation{\oldtheequation\alph{equation}}}

\def\endsubequations{%
  \setcounter{equation}{\@savedequation}%
  \@stequation=\expandafter{\@savedtheequation}%
  \edef\theequation{\the\@stequation}%
  \global\@ignoretrue}

\def\big#1{{\hbox{$\left#1\vcenter to1.428\ht\strutbox{}\right.\n@space$}}}
\def\Big#1{{\hbox{$\left#1\vcenter to2.142\ht\strutbox{}\right.\n@space$}}}
\def\bigg#1{{\hbox{$\left#1\vcenter to2.857\ht\strutbox{}\right.\n@space$}}}
\def\Bigg#1{{\hbox{$\left#1\vcenter to3.571\ht\strutbox{}\right.\n@space$}}}

\lock

\def\uone{$U(1)$}
\def\utwo{$U(2)$}
\def\un{$U(N)$}
\def\son{$SO(N)$}

\def\sotwo{$SO(2)$}
\def\on{$O(N)$}
\def\ozero{$O(0)$}
\def\oone{$O(1)$}
\def\otwo{$O(2)$}

\def\A#1#2{A_{\p1 #1}{}^{#2}}
\def\Aij{\A i j}
\def\AIJ{A_{I J}}

\def\xsp{\p1}
\def\alphastar{\alpha^{*}}
\def\psistar{\psi^{*}}
\def\chistar{\chi^{*}}

\def\Xstaro#1{X^{* \xsp #1}}
\def\chistaro#1{\chi^{* \xsp #1}}
\def\alphastaro#1{\alpha^{* \xsp #1}}
\def\psistaro#1{\psi^{* \xsp #1}}
\def\chistart#1#2{\chi^{* \xsp #1 \xsp #2}}
\def\chidown#1#2{\chi_{#1 \xsp #2}}

\def\liststuff#1{\\\penalty -500 $\bullet$ {\bf #1}\\*}

\footnotesections

\begin{document}
\begin{titlepage}

\noindent September 28, 1994\hfill    TAUP--2199--94\\
\null\hfill          hep-th/9409175

\vskip 1.5 cm

\begin{center}

{\large \bf
\Kahler{} spinning particles
}

\vskip 1 cm
{
Neil Marcus\footnote{
Work supported in part by the US-Israel Binational Science Foundation,
the German-Israeli Foundation for Scientific Research and Development
and the Israel Academy of Science.\\
E--Mail: NEIL@HALO.TAU.AC.IL}
}
\vskip 0.3 cm

{\sl
School of Physics and Astronomy\\Raymond and Beverly Sackler Faculty
of Exact Sciences\\Tel-Aviv University\\Ramat Aviv, Tel-Aviv 69978, ISRAEL.
}

\end{center}

\begin{abstract}

We construct the \un{} spinning particle theories, which describe particles
moving on \Kahler{} spaces.  These particles have the same relation to the
$N=2$ string as usual spinning particles have to the NSR string.  We find the
restrictions on the target space of the theories coming from supersymmetry
and from global anomalies.  Finally, we show that the partition functions of
the theories agree with what is expected from their spectra, unlike that of
the $N=2$ string in which there is an anomalous dependence on the proper
time.

\end{abstract}

\end{titlepage}

\newpage

\section{Introduction}

The ``spinning particle'' \cite{spart} describes a free Dirac
particle, moving in some $D$--dimensional space.  Historically, the
particle action led to that of the NSR string.  Conversely,
one can obtain the spinning particle by dimensionally
reducing the NSR string \cite{sstring} or the heterotic string
\cite{heter} to one dimension.  The particle can be generalized to an
$N$--spinning particle with a gauged \on{} symmetry, which in four
dimensions describes a spin $N/2$ particle
\cite{russian\on,\on}.  The string can also be generalized, to the
$N=2$ \cite{n=2string} and $N=4$ \cite{quater} strings.  However, the
dimensional reduction of these extended string
theories does {\em not\/} give the \on{} spinning particles.  The
(ungauged) \otwo{} particle can be obtained by the dimensional
reduction of the NSR string, but the $N>2$ theories can not be derived
from string theories.

In this paper we shall construct the \un{} spinning particles, which have the
same relation to the $N=2$ string as the \on{} particles have to the NSR
string.  As with the $N=2$ string, these theories are not directly relevant
to the real world, since they always have an even number of time coordinates.
Our original motivation for studying them was that since particle theories
are so much simpler than string theories, the \un{} particles could provide us
with an insight into some of the puzzles posed by $N=2$ strings.  These
include the Lorentz-invariance and supersymmetry of the string
\cite{allWarren}, and the conflict between loop calculations in the string
and in the corresponding field-theories \cite{MM,italians}.  (For a review
see \cite{italy}.)

A separate motivation is that spinning particle theories are
one-dimensional supergravity theories\footnote{This is not the
most general definition of a particle theory.  For example, one has the
superparticle~\cite{superpart} which is invariant under target-space
instead of world line supersymmetries, as well as hybrid
theories~\cite{sspinpart} that are
combinations of the two.  We shall not consider such theories further.},
and since most supergravity theories turn out to be
useful in one way or another, they are interesting to consider in their own
right.   Thus, truncations of the \uone{} and \utwo{}
theories of this paper have already been used to
give a particle description of the open and closed
B--twisted topological sigma models, respectively \cite{Bpart}.

In this regard,
it is useful to recall some results in the classification of supergravity
theories: first, we should perhaps stress the difference between
supergravity theories in three to eleven dimensions, and those in one and
two dimensions.  Supergravity theories in $D>2$ are related by
dimensional reduction and truncation.  The most beautiful---although
possibly the least useful---ones are the ``larger'' supergravities ($N>2$
in four dimensions).  They are essentially unique, with their
scalar fields living in various homogeneous spaces \cite{5d}.  (References on
supergravity theories in various dimensions can be found in
\cite{various}.)  The smaller theories are more complicated, since matter
supermultiplets can be coupled to them in various ways.  Thus $N=1$ theories
in three dimensions can be written with the scalar fields describing a sigma
model on an arbitrary
Riemann space \cite{3d}.  The scalars of $N=1$ theories in four dimensions
describe a Hodge manifold \cite{BW},
while $N=2$ theories in four dimensions lead to a quaternionic sigma model
\cite{BW}.  The supergravity theories
can exist on any spacetime, as long as it is a spin manifold.

Two-dimensional supergravity theories are not the dimensional reduction of
those in three dimensions.  While it is not necessary to do so, their main
interpretation is as string theories, with the scalar fields
interpreted as coordinates on a target space which is spacetime.  Thus
one might expect the larger ($N>4$) supergravity theories to live on particular
spacetimes.  Even if these theories exist, they would be
rather esoteric, and they have not been constructed.
Classically $N=1$, 2 and 4 strings live on Riemann \cite{salam}, \Kahler{}
\cite{quater} and either hyper\kahler{} or quaternionic \cite{quater}
spaces\footnote{Of course demanding conformal invariance of the string theory
restricts the spaces to be essentially Ricci flat, and phenomenological
constraints may lead one to further restrict the theories, for example
to \CY{} spaces.}.  (Recall that global
$N=1$, 2 and 4 sigma models live on Riemann \cite{freedman}, \Kahler{}
\cite{22} and hyper\kahler{} \cite{AGF} spaces, respectively.)  In
addition, one also has the various
heterotic strings.  This certainly is not a general classification of string
theories---for example one can introduce torsions into the string---but
it does give an overview of the basic types of string theories.

The two-dimensional theories {\em can\/} be reduced to one dimension, where
they become particle theories.  The reduction of the NSR
string can be generalized to give the usual \on{} spinning particles, which
exist on Riemann spaces (sometimes on spin manifolds only).  The \un{}
theories to be considered here can be derived from the $N=2$ string, and they
can be defined only on \Kahler{} target spaces.  One can also
compactify the $N=4$ string, which we expect to give $USp(N)$ spinning
particles living on hyper\kahler{} or quaternionic spaces; however we
shall not consider these theories further in this work.

The rest of this paper is organized as follows: In section two we
give a brief summary of the known spinning particle actions, in order
to compare them to the \Kahler{} spinning particles.  Most of this section
is a restatement of results in \cite{\on} in our notation.  In section
three we construct the \un{} theories.  We discuss the restrictions on the
target space of the theories coming from supersymmetry and, after introducing
a \CS{} term, from anomalies.  We then find the spectra of the
theories, and discuss their space-time
conformal invariance.  In section four
we calculate the one-loop partition function of the particle, and see
that it gives the result expected from its spectrum, unlike the
corresponding calculation  in the string.  We end with some conclusions.

\section{Summary of the \on{} spinning particle}

The simplest particle action is simply that of an (unspinning) scalar
particle with mass $m$ moving in a $D$--dimensional Minkowski space.  It
can be described by the action \cite{polybook}:
\beq
\call = \, \frac1{2e} \; \dot X ^M \dot X ^N + \frac e2 \; m^2 \comma
\eeq
where $X^M$ is a map from the world line of the particle to the target
space, and $e$ is an \einbein{} on the world line.  Canonical quantization
shows that $X^M$ and $P^N = \dot X^N/e$ have the usual commutation
relations of coordinates and momenta, and the equation of motion of the
\einbein{} gives the constraint $P^2 = m^2$.

The spinning particle has a one-dimensional supersymmetry, which is made
local by the introduction of a gravitino $\psi$.  The supersymmetric
partners of the $X^M$'s are the spinors $\chi^M$'s, and the action in the
massless case is given by \cite{spart}:
\beq
\call = \, \frac1{2e} \; \dot X^M \dot X ^N
                + \frac i e \dot X^M \, \psi \, \chi^M
                + \frac i2 \; \chi^M \dot \chi^M
\stop \label{Dirac}
\eeq
Canonically quantizing \noeq{Dirac}, one sees that the $\chi^M$'s become
gamma matrices, so one is describing a Dirac particle\footnote{In the
massive case, one needs to introduce an extra spinor which becomes
$\gamma^5$ upon quantization \cite{spart}.}.  The importance of the local
supersymmetry of the action is seen from the fact that the constraint
coming from the equation of motion of the gravitino is the massless Dirac
equation.

This construction can be generalized to the $N=2$ case
\cite{spart2,russian\on,\on},
where one has two gravitini $\psi_I$ and $2D$ spinors $\chi_I^M$.  This
theory has a gauged \sotwo{} symmetry, and in four dimensions it describes
the field equations of a Maxwell field.  One can continue generalizing to the
$N$--extended spinning particle \cite{russian\on,extra\on,\on}, with gravitini
$\psi_I$ and spinors $\chi_I^M$ in the $N$ of a local \on{}.  The lagrangian
of the massless ``\on{} spinning particle'' in a $D$--dimensional Riemann space
is \cite{\on}:
\beqal
\call = \, \frac1{2e} \; G_{M N} \,
           & \left( \, \dot X^M
                                + i \, \psi \cdot  \chi^M \, \right)
           \left( \, \dot X ^N + i \, \psi \cdot \chi^N \, \right)
        + \frac i2 \; G_{M N} \; \chi_I^M \, \cald \, \chi_I^N \\
        & - \, \nfrac{e}8\,
        R_{M N P Q} \; \chi^M \cdot \chi^N \; \chi^P \cdot \chi^Q
\comma \label{O(N) particle}
\eeqal
where the ``dots'' denote contractions over \on{} indices.
Here $\cald$ is the covariantized time derivative, improved with a connection
for the
\on{} group and the pullback of the Christoffel connection:
\beq
\cald \, \chi_I^M \equiv \dot \chi_I^M - i \, \AIJ \, \chi_J^M +
        \Gamma_{P Q}^M \, \dot X ^P \chi_I^Q \stop \label{O(N) cov}
\eeq

At this stage the metric $G_{M N}$ appears to be arbitrary,
but one can see that for $N>2$
supersymmetry forces the theory to be in flat space.  Using the Noether
procedure, one finds the local supersymmetry transformations:
\beqaltwo
&\delta \, e =  -2 \, i \, \alpha \cdot \psi
&&\delta \, \AIJ  = 0 \\
&\delta \, \psi_I = \cald \, \alpha_I
&&  \yesnumber \label{O(N) susy} \\
&\delta \, X^M =  -i \, \alpha \cdot \chi^M
&&\delta \, \chi_I^M = \nfrac1e
        \left( \, \dot X^M + i \, \psi \cdot \chi^M \, \right) \alpha_I
                + i \, \Gamma_{P Q}^M \; \alpha \cdot  \chi^P  \chi_I^Q \stop
\eeqaltwo
Under these, the lagrangian is invariant up to 3--fermi terms (and total
derivatives), but one is left with the 5--fermion terms:
\beqal
\delta \call = \frac i4 \;  &R_{M N P Q} \,
        \left( \alpha \cdot \psi \chi^M \cdot \chi^N -
           2 \, \psi \cdot \chi^M \alpha \cdot \chi^N \right)
                \chi^P \cdot \chi^Q \\
     + & \, \frac {i\,e}8 \;  R_{M N P Q \xsp ; \xsp R}  \;
                \chi^M \cdot \chi^N \chi^P \cdot \chi^Q \alpha \cdot \chi^R
                \stop \label{O(N) extra}
\eeqal
For the Dirac particle ($N=1$) these extra terms vanish due to the
symmetries of the Riemann tensor, so the action is supersymmetric.  (In
fact, in this case both of the 4-fermi terms in the lagrangian
\noeq{O(N) particle} vanish identically.)  Similarly, for $N=2$ the
5--fermion terms vanish using the symmetries of the Riemann tensor and the
Bianchi identity \cite{\on}.  However, for $N>2$ the lagrangian is
supersymmetric only for a flat target space, with $R_{M N P Q}=0$, so the
$O(N>2)$ spinning particle is relatively uninteresting.

Without going into further detail, the \on{} spinning particle has the
following properties:
\liststuff{Spectrum}
The ``\ozero{}'' theory describes a scalar moving
in any Riemann space.  The ``\oone'' theory describes a Dirac spinor.
This theory has a global anomaly unless the target space is a spin
manifold \cite{witanom,AG}.  The \otwo{} theory describes an antisymmetric
tensor field with $D/2$ indices---a photon if $D=4$.  The theory has a
global anomaly in odd dimensions.
The $O(N>2)$ theory can be written only in
flat space.  In four dimensions it describes a spin $N/2$ particle
\cite{russian\on,\on}.  In $D$ (even) dimensions, it describes a particle whose
representation is described by the rectangular Young tableaux with $D/2$
rows and $N/2$ columns \cite{confpart,\on}, with half a column representing a
spinor index.
\liststuff{\CS{} term}
In the case $N=2$, the gauge group of the theory is an $SO(2) \simeq U(1)$.
Thus, one can add the term $\epsilon^{IJ} A_{IJ} = \Tr A$ to the
lagrangian.  This is
the simplest example of a \CS{} term.  Note that this term breaks the \otwo{}
of the theory to an \sotwo.  With the
addition of the \CS{} term with coefficient $q-D/2$, the theory describes an
antisymmetric $q$--tensor field in any $D$--dimensional Riemann space.
($D$ can now be arbitrary, but $q$ must be an integer to avoid the global
anomaly \cite{\on}.)  Thus the $N=2$
theory can describe any antisymmetric tensor particle on any Riemann space.
\liststuff{Conformal invariance}
The massless \on{} theory is
invariant under target-space dilations.  In fact the theory is even invariant
under conformal transformations of the target-space \cite{confpart,\on}, and
all conformal representations in all dimensions can be obtained from the
\on{} theory \cite{allconf}.
In the \sotwo{} theory conformal invariance is spoiled if the
\CS{} term is added, and indeed the theory of a general
antisymmetric tensor field in $D$ dimensions is not conformally invariant.
\liststuff{Supersymmetry algebra}
The supersymmetry algebra  in the \on{} theory closes into field-dependent
diffeomorphisms, supersymmetry transformations and gauge transformations.
If $N>1$, one has to use the fermion equations of motion, so
the algebra closes only on shell.

Finally, note that the gauge field $A_{IJ}$ does not transform under
the supersymmetry transformations in \noeq{O(N) susy}.  This means that if
one's interest is in writing the most general one-dimensional supergravity
theory---rather than a particle theory---one
is free to gauge any subgroup of the \on{} symmetry, from
the full \on{} to the trivial identity group.  This is unlike the case of
the string or of
supergravity in higher dimensions.  If one does not gauge the full
group, the theory will not describe a particle in an irreducible
representation of the Lorentz (or conformal) group.  For example, if one
does not introduce the gauge field in the \otwo{} theory, the theory
describes all antisymmetric tensor fields simultaneously.

\section{The \un{} spinning particle}
\subsection{Lagrangian}

We have argued that the \on{} particles of the previous section are related
to the NSR string.  To carry out the dimensional reduction, one first fixes
the Weyl, super-Weyl and Lorentz transformations, as well as the spatial
diffeomorphisms of the string by the gauge choices $e_\alpha^a = \diag
\left( e ,1 \right)$ and $\psi_1=0$.  Then one gets the (ungauged)
\otwo{} particle, which can be truncated to the usual spinning
particle\footnote{One might have expected to have obtained the
\otwo{} gauge field from the $e_0^1$ component of the \zweibein, but this
drops out of the action.}.  The \otwo{} particle is {\em not\/}
related to the $N=2$ string, as one might have expected from its \uone{} gauge
symmetry.  This follows simply by counting fields:  In the reduction
of the $N=2$ theory, there are twice as many spinors as scalars, as in the
\otwo{} particle, but there are 2 {\em complex\/} instead of 2 real gravitini.

Thus reducing the $N=2$ string to one dimension leads to a new family of
particle theories.  The $N=2$ string lives on a $d$--complex dimensional
\Kahler{} space ($D=2d$), so the
coordinates $X^M$ are split into $X^\mu$ and their complex conjugates
$\Xstaro \mubar$.  In analogy to the \son{} particle, we introduce an index
$i$ which will transform under a local \un.  Then the spinors become
$\chi_i^\mu$ and $\chistart i\mubar$, and the gravitini become $\psi_i$ and
$\psistaro i$.  The lagrangian can be found by reducing the
$N=2$ string (in the \utwo{} case with only a \uone{} gauging), or simply
by using the Noether procedure:
\beqal
\call = \, \frac1e \; G_{\mu\mubar} \,
           & \left( \, \dot \Xstaro \mubar
                                + i \, \psi \cdot \chistaro \mubar \, \right)
           \left( \, \dot X ^\mu + i \, \psistar \cdot \chi^\mu \, \right)
        + i \, G_{\mu\mubar} \, \chistart i\mubar \, \cald \, \chi_i^\mu \\
        & - \, \nfrac{e}2\,
        R_{\mu\mubar\xsp\nu\nubar} \, \chi^\mu \cdot \chistaro \mubar \,
                                        \chi^\nu \cdot \chistaro \nubar
\stop \label{particle}
\eeqal
Here the
derivative $\cald$ is again covariantized with respect to diffeomorphisms
and with respect to the local \un{}:
\beq
\cald \, \chi_i^\mu \equiv \dot \chi_i^\mu - i \, \Aij \, \chi_j^\mu +
        \Gamma_{\rho\sigma}^\mu \dot X ^\rho \chi_i^\sigma \stop \label{cov}
\eeq
Recall that in a \Kahler{} space, the only nonvanishing components of the
Christoffel symbol are $\Gamma_{\mu\nu}^\rho$ and its complex conjugate.

\subsection{Supersymmetry}
The supersymmetry transformations are given by
\beqal
&\delta \, e =  -i \, \alpha \cdot \psistar \,
                -i \,  \alphastar \cdot \psi  \\
&\delta \, \psi_i = \cald \, \alpha_i \\
&\delta \, \Aij  = 0 \\
&\delta \, \Xstaro \mubar = -i \, \alpha \cdot \chistaro \mubar \\
&\delta \, \chidown i\mubar = \nfrac1e \, G_{\mu\mubar}
                \left( \, \dot X^\mu + i \, \psistar \cdot \chi^\mu \, \right)
                \alpha_i
                - i \, \Gamma_{\mubar\rhobar,\sigma} \;
                \alpha \cdot \chistaro \rhobar \, \chi_i^\sigma
                        \stop \label{susy} \\
\eeqal
We have chosen to write the transformation of $\chidown i\mubar$ with its
spacetime index lowered, since the transformation
of $\chi_i^\mu$ involves both $\alpha$ and $\alphastar$.  As in the
\on{} case, one finds 5--fermi terms in the supersymmetry variation of the
lagrangian:
\beqal
\delta \call = \frac i2 \;  &R_{\mu \mubar \nu \nubar} \,
      \left( \alpha \cdot \psistar \, \chi^\mu \cdot \chistaro \mubar -
         2 \, \psistar \cdot \chi^\mu \, \alpha \cdot \chistaro \mubar \right)
                \chi^\nu \cdot \chistaro \nubar \\
     + & \, \frac {i\,e}2 \;  R_{\mu \mubar \nu \nubar \xsp ; \xsp \rhobar} \;
                \chi^\mu \cdot \chistaro \mubar \,
                \chi^\nu \cdot \chistaro \nubar \,
                \alpha \cdot \chistaro\rhobar \, + \, \hbox{h.c.} \label{extra}
\eeqal
In the \uone{} theory the curvature terms vanish both here
and in the lagrangian \noeq{particle},
since in a \Kahler{} space $R_{\mu \mubar \nu \nubar}$ is symmetric in each
set of indices.  In the \utwo{} theory the extra terms
in \noeq{extra} vanish because of the symmetries of
$R_{\mu \mubar \nu \nubar}$ and because of the Bianchi identity, which in a
\Kahler{} space states that
$R_{\mu \mubar \nu \nubar \xsp ; \xsp \rhobar}$ is totally symmetric in the
``barred'' indices.  Thus the \uone{} and \utwo{} theories can be written in
any \Kahler{} space.  As in the \on{} case, the $U(N>2)$ theories are
supersymmetric only in flat space, and are again of limited interest.

Note that since $\Aij$ does not transform under supersymmetry
transformations, one can again restrict the gauging of the theory to any
subgroup of the \un{} symmetry.  For example, the open and closed
$B$--particles of ref.~\cite{Bpart} are given by the \uone{} and \utwo{}
lagrangians of \noeq{particle}, with no gauging whatsoever and with the
$\psi_i$'s set to zero.  (Thus in these theories only the $\alphastaro i$
supersymmetry transformations are local; those of the $\alpha_i$'s are
not.)

Finally, upon commuting two supersymmetry transformations with parameters
$\alpha_i$ and $\beta_j$, one finds a diffeomorphism by
$\xi \equiv i/e \left( \alphastar \cdot \beta + \alpha \cdot \beta^* \right)$,
a supersymmetry transformation with parameter $- \xi \, \psi_i$ and a \un{}
transformation with parameter $-\xi \, \Aij$.  For $N>1$, the algebra
closes only with the use of the equation of motion of the $\chi_i^\mu$'s.

\subsection{\CS{} term, the spectrum, and anomalies.}

The spectrum of the particle can be changed by adding the \CS{} term
\beq
\call_{CS} =
\left( \frac d 2 - q \right) \, \A i i \label{\CS}
\eeq
to the lagrangian of the particle, where for now $q$ is an arbitrary
parameter.  Note that the \CS{} term exists for all $N$,  and that it
is consistent with supersymmetry, since $\Aij$ is totally invariant.

In canonically quantizing the theory, the $X^\mu$'s and their conjugates
$P_\mu$ become position and momentum operators, as do $\Xstaro \mubar$ and
$P_\mubar^*$.   The $\chistart i \mubar$'s can be taken to be creation
operators and the $\chi_i^\mu$'s to be annihilation operators, so
the general state with momentum $P_\mu$ is built from the vacuum state
$\ket{P_\mu}$ by applying some number of $\chistart i \mubar$'s.  As usual, the
equations of motion of the supergravity fields $e$, $\psi_i$ and $\Aij$
constrain the Hilbert space: Thus, varying $\Aij$ in the combined lagrangian
of \noeq{particle} and \noeq{\CS} gives:
\beq
G_{\mu\mubar} \, \chistart i \mubar \, \chi_j^\mu = q \; \delta_j^i
        \comma\label{constr}
\eeq
where we have used a normal ordering scheme that is symmetric between the
$\chi$'s and $\chi^*$'s.  Acting the ``$ii$'' element of this constraint on a
state $\Psi$ tells us that there must be exactly $q$ $\chistaro i$'s in the
state for each $i$, so $\Psi$ has the form:
\beq
\Psi = F_{\mubar_1 \cdots \mubar_q \; \nubar_1 \cdots \nubar_q \; \cdots}
       \; \chistart 1 {\mubar_1} \cdots \chistart 1 {\mubar_q}
       \; \chistart 2 {\nubar_1} \cdots \chistart 2 {\nubar_q}
       \; \cdots \; \ket{P_\mu} \label{Psi} \stop
\eeq
This means that the theory is empty unless $q$ is an integer between
0 and $d$.  (Thus the \CS{} term is {\em necessary\/} in an odd number of
complex dimensions).  The lack of a spectrum when $q$ is not an integer is an
indication of the global anomaly of the theory in that case
\cite{israelis}.

The off-diagonal elements of the constraint \noeq{constr}
impose a symmetry between the $i$ and $j$ indices of the tensor $F$,
implying that it is represented by the rectangular Young tableaux
\beq
q\left\{\vbox to 4.25ex{} \right. \n5
         \underbrace{\boxup4{\boxes5\boxes5\boxes5}}_N \label{q boxes}
\eeq
of $SU(d)$.  (This is similar to the case of the \on{} particle, where
the state has $D/2$ rows and $N/2$ columns.)
\ju{Note that only the $SU(d)$ representation of the state $\Psi$ is
explicitly seen in \noeq{Psi}.}
In general, the holonomy group of the
\Kahler{} space will be a full $U(d)$.  Again using a symmetric normal
ordering, and normalizing the \uone{} charge of
$\chi_i^\mu$ to be 1, one sees that the particle represented by $\Psi$ has
charge \hbox{$N\,(d/2-q)$}.  If $N \, d$ is odd the particle will have a
half-integral \uone{} charge, indicating that it is spinor-like.  (The simplest
case with $N=1$ gives the various pieces of the $SO(D)$ spinors broken into
$U(d)$ representations, as $q$ is varied.)  In these cases the theory
again has a global anomaly unless the space
supports a spin structure.  In a \CY{} space this problem never arises,
since the holonomy group of the spacetime is $SU(d)$.

The equations of motion of the state $\Psi$ are given by varying the
lagrangian with respect to the gravitini and the \einbein.  In order to
avoid normal ordering problems we shall for simplicity restrict ourselves
here to the case of flat space\footnote{Such problems lead to an unusual
choice of creation and annihilation operators in the
B--particle~\cite{Bpart}.}.  The \einbein{} constraint shows that the
particle is massless: $G^{\mu\mubar} P_\mu P_\mubar^* = 0$, while the
constraints from the gravitini give the equations of motion
\beqal
&  P^{\mubar_1}
  F_{\mubar_1 \cdots \mubar_q \xsp\nubar_1 \cdots \nubar_q \xsp \cdots} = 0 \\
&  P_{\,[\xsp \mubar}^* \,
  F_{\mubar_1 \cdots \mubar_q \xsp ] \xsp\nubar_1 \cdots \nubar_q \xsp \cdots}
 = 0 \label{Dirac eqs} \stop
\eeqal
Eqs\eq{Dirac eqs} are analogous to the equation of motion and Bianchi
identity of the photon (which one can get from the \otwo{} theory):
$\pa_M \, F_{MN} = 0$ and $\pa_{\,[\xsp L} \, F_{MN \xsp]}=0$.  By going into a
light-cone frame, one can see that the field strength in \noeq{q boxes} is
reduced on shell to a connection in the
\beq
q-1 \left\{\vbox to 3.8ex{} \right. \n5
       \underbrace{\boxup3{\boxes4\boxes4\boxes4}}_N
                \label{q-1 boxes}
\eeq
representation of the massless little group $U(d-2)$.
Note that the theory has no propagating particles for $q=0$ and $q=d$.

\subsection{Conformal invariance}

In addition to the worldline symmetries we have discussed, the \un{}
particle is covariant under holomorphic diffeomorphisms.  It is therefore
invariant under holomorphic isometries of the spacetime.  For example, in
flat space it is invariant under $U(d)$ ``Lorentz'' transformations.  The
bosonic ($N=0$) string is invariant under the more general class of
conformal transformations, defined by $\delta \, X^\mu = \xi^\mu$, with
$\xi^\mu$ restricted to be a conformal Killing vector:
\begin{subequations}
\begin{eqalignno}
&   \xi_{\mubar \xsp , \xsp \mu} + \xibar_{\mu \xsp , \xsp \mubar} =
     2 \, G_{\mu \mubar} \, \rho  \label{scale} \comma \\
& \xi_{\mubar \xsp ; \xsp \nubar} +
        \xi_{\nubar \xsp ; \xsp \mubar} = 0 \label{oconf} \semi
\end{eqalignno}
here the ``scale-factor'' $\rho$ defined by \noeq{scale} vanishes if
$\xi^\mu$ is a Killing vector, in which case one is back to an isometry
transformation of the spacetime.  If one attempts to generalize the
conformal symmetry to the \un{} lagrangian, one finds that a term
$$
\delta \call = \frac i e \; \xi_{,\xsp \nubar}^\mu \,
\dot \Xstaro \nubar \, \psi \cdot \chistar_\mu \, + \, \hbox{h.c.}
$$
can not be canceled.  This
means that \noeq{oconf} must be replaced by the stronger condition that
$\xi^\mu$ be a {\em holomorphic\/} conformal Killing vector, with
\beq
\xi_{,\xsp \nubar}^\mu =0 \label{nconf} \stop
\eeq
\end{subequations}

In addition, one finds the unwanted terms
$$
\delta \call = i \, \dot \Xstaro \mubar \, \chistaro \nubar \cdot \chi^\mu
          \left( -\rho_{,\xsp \mubar} \, G_{\mu\nubar} +
          \rho_{,\xsp \nubar} \, G_{\mu\mubar} \right) \, + \, \hbox{h.c.}
$$
These terms vanish identically when $d=1$, but otherwise one must impose the
condition that $\rho$ be constant.  Aside from holomorphic isometries of the
spacetime, this leaves only constant dilations.  By going to
Riemann-normal coordinates one can see that such
holomorphic dilations are only possible in flat space.
In this case there is clearly a dilation symmetry, with the fields
$X^\mu$, $e$, $\chi_i^\mu$,
$\psi_i$ and $\Aij$ having weights 1, 2, 0, 1 and 0, respectively.

We thus see that there are only interesting conformal transformations when
the space-time is a Riemann surface $(d=1)$.  (Note, however, that Riemann
surfaces can not support massless propagating particles, since they can not
have both space and time coordinates, so the theory is basically topological
in this case.)  Then one does have invariance under the full conformal
group, with the transformations:
\beqal
&\delta \, X^\mu =  \xi^\mu \\
&\delta \, \chi_i^\mu = \xi^\mu{}_{,\xsp\nu} \, \chi_i^\nu
                                - \rho \, \chi_i^\mu \\
&\delta \, e = 2 \, \rho \, e \yesnumber \label{conf} \\
&\delta \, \psi_i = \rho  \, \psi_i  + e \rho_{,\xsp\mu} \, \chi_i^\mu \\
&\delta \, \Aij = \frac1{N+1} \,
                \left( \delta_i^j \delta_l^k - N \delta_i^k \delta_l^j \right)
            \left(\rho_{,\xsp\mu} \, \psistaro l \chi_k^\mu
              - \rho_{,\xsp\mubar} \, \psi_k \chistart l \mubar
              - e \, \rho_{,\xsp\mu\mubar} \, \chistart l \mubar
                        \chi_k^\mu  \right) \stop
\eeqal
The transformation of $\Aij$ can actually be written as various linear
combinations of the two sets of terms in \noeq{conf}; we have chosen the
combination that leaves $\Tr A$ invariant, so the \CS{} lagrangian
\noeq{\CS} does not break the conformal symmetry.

\section{The partition function}

Thus far, we have found the spectrum of the particle using Hamiltonian
quantization.  It would be nice to also be able to calculate amplitudes of the
theory.  However, since the particle lagrangian describes a free particle, the
only quantity one can calculate without introducing interactions
is the one-loop partition function.  This is nevertheless
interesting, since the analogous
partition function of the $N=2$ string disagrees with the result expected
from its spectrum.  Recall that the partition function of a particle with
mass $m$ moving in a Euclidean $D$--dimensional space is \cite{Polchinski}:
\beqal
\calf & = - \, \frac12 \; \Tr \,\log \, \left( p^2+m^2 \right) \\
      & = \, \frac V2 \int_0^\infty  \frac {{\rm d}T} {T^{\, 1+D/2}}
      \; e^{-m^2 \, T } \comma \label{part}
\eeqal
where $T$ is the Schwinger proper-time parameter.
The $N=2$ string describes only massless degrees of freedom in a flat
2--{\em complex\/} dimensional space ($d=2$, $D=4$).  However, its partition
function is \cite{MM}
\beq
\calf = \frac12 \, \int_\calm {{\rm d}^2\tau \over \tau_2^2}
     \comma \label{string}
\eeq
which is compatible with that of the particle only if $D=2$!  There is a
similar disagreement between the one-loop three-point function in the string
\cite{italians} and that of the ``Plebanski'' field theory
\cite{Vafa N=2 closed} which should be the effective field theory describing
the string.
One of the explanations advocated to solve this problem  \cite{Vafa N=2 closed}
is that the complex nature of the \Kahler{} target space of the $N=2$ string
means that one should use a complex Schwinger parameter.
The problem could also come from the \uone{} gauge field in the theory, or
have an intrinsically stringy nature.  Since the \un{} particle
is intimately related to the $N=2$ string,  and shares many of its features,
calculating its partition function can distinguish between the various
possibilities.

We shall start with the \uone{} theory in flat space
with complex dimension $d$, on a worldline which is a circle with proper
length $T$.  We need to evaluate the path integral over the lagrangian
\noeq{particle} with the \CS{} term \noeq{\CS}.  The new feature of the \un{}
string is that the gauge field $A$ can not be completely gauged away, since
it can have a nontrivial Wilson line around the circle.  The allowed
gauge choice is $A = \theta/T$, with $\theta$ ranging from 0 to $2 \pi$.
This means that the spinors
$\chi^\mu$ and the gravitino $\psi$ pick up a phase of $\theta$, when going
around the circle.  Most of the path integral is standard:
As usual \cite{Polchinski,polybook}, the integral over the \einbein{} (gauged
to $e=T$) modulo diffeomorphisms gives
\beq
\frac12 \int_0^\infty  \frac {{\rm d}T} T \comma \label{e}
\eeq
and the path integral over the $X^\mu$'s gives
\beq
V T^d \, \det' \bigl( -\partial_t{}^2 \bigr)^{-d} = \frac V {T^d}
                 \stop \label{X}
\eeq
Similarly, the integral over the  $\chi^\mu$'s gives a factor of
\beq
\bigl( \det_{\theta} \, i \partial_t \bigr)^d \comma \label{chi}
\eeq
where the index on the ``det'' is to remind us of the shifted boundary
conditions of the fermions.  For a generic $A$, the gravitino can be
completely gauged away using the supersymmetry transformation, with the
resulting Jacobian reducing the exponent in \noeq{chi} to $d-2$.

The interesting question, of course, is what comes from the integral over
$A$, modulo the \uone{} transformations.  The zero-mode integration gives
$\int_0^{2\pi} {\rm d} \theta /\sqrt{T}$; the integration over constant
gauge transformations gives a factor of $2\pi \sqrt{T}$ in the denominator,
and the Jacobian is $\det' \, i \partial_t =T$.  All together, this means
that the $A$ integration simply gives rise to an insertion of the projection
operator $\int_0^{2\pi} \frac {{\rm d}\theta} {2\pi}$.  Since the gauge-field
integration contributes no factors of $T$, there will be no
anomalous powers of $T$ in the particle.

Putting together the above results, and
adding the \CS{} term \noeq{\CS}, we find
\beq
\calf = \, \frac V2 \int_0^\infty  \frac {{\rm d}T} {T^{\, 1+d}}
        \; \int_0^{2\pi}  \frac {{\rm d}\theta} {2\pi}
        \; e^{i\, (q-d/2) \, \theta}
        \; \bigl( 2 \sin \theta/2 \bigr)^{d-2}
        \comma \label{\uone}
\eeq
where we have used the result that (up to an ill-defined sign)
$\det_{\theta} \, i \partial_t = 2 \, \sin ( \theta/2 )$.
Evaluating the $\theta$ integral, and  comparing \noeq{\uone}
to \noeq{part}, one sees that the \uone{} particle describes
$d-2 \choose q-1$ massless particles in a $d$--complex dimensional space.
This is exactly the number of degrees of freedom that one expects from the
Hamiltonian calculation: the field strength of the particle is described by
an antisymmetric tensor with $q$ indices \noeq{q boxes}; the equations of
motion \noeq{Dirac eqs} reduce this to a light-cone connection with $q-1$
indices \noeq{q-1 boxes}.

To check whether or not subtleties occur when the gauge group is
nonabelian, we shall also carry out the calculation of the partition function
in the \utwo{} theory.  (The combinatorics of the \un{} case will be left to
our more intrepid readers.)  The integrals over $e$ and $X$ are the same as
before.  In this case $\Aij$ can be gauge fixed to $A = \diag \left( \theta ,
\phi \right)$, where both $\theta$ and $\phi$ go from 0 to $2 \pi$.  The
fermionic integrations are the same as in the \uone{} result
\noeq{chi}, duplicated over $\theta$ and $\phi$.
Similarly, the integrations over the diagonal zero modes of $A$
give projection operators for both $\theta$ and $\phi$.  The only new
features are that there is a factor of $1/2$, since a $90^\circ$ rotation in
the $y$--$z$ plane interchanges $\theta$ and $\phi$, and that the
two off-diagonal gauge transformations each
contribute the Jacobian $\det_{\theta-\phi} \, i \partial_t$.  The
integrations over $\theta$ and $\phi$ then give a factor of
\beqal
 -\,\frac 12 & \int_0^{2\pi}  \frac {{\rm d}\theta} {2\pi}
         \int_0^{2\pi}  \frac {{\rm d}\phi} {2\pi}
        \; e^{i\, (q-d/2) \, (\theta+\phi)}
        \; \bigl( 2 \sin (\theta/2-\phi/2) \bigr)^2
        \; \bigl( 2 \sin \theta/2 \bigr)^{d-2}
        \; \bigl( 2 \sin \phi/2 \bigr)^{d-2} \\
        &= \; \frac{(d-1)!\, (d-2)!}{(d-q)!\, (d-q-1)!\, q!\,(q-1)!} \stop
\eeqal
This is the dimension of the $U(d-2)$ representation of the
particle in \noeq{q-1 boxes}, for $N=2$, so the path integral
calculation of the partition function is again in perfect agreement with the
result expected from the canonical quantization of the particle.  There is
no sign of the anomalies of the $N=2$ string.

\section{Conclusions}

We have constructed the \un{} spinning particles---one dimensional
supergravity theories with $N$ complex local supersymmetries on the world
line and  a local \un{} invariance.
These theories describe massless particles moving in a \Kahler{}
spacetime with complex dimension $d$.  The \uone{} and \utwo{} theories can
be defined on any \Kahler{} space; consistency with supersymmetry forces the
$U(N>2)$ theories to be in flat space.  The theories have a spacetime conformal
invariance only if the target space is a Riemann surface ($d=1$).

The spectrum of the theories depends on the coefficient of the \CS{} term
$(d/2-q) \; \Tr A$, which can be added to the lagrangian for any $N$.  To avoid
global anomalies $q$ must be an integer, and the manifold must support a spin
structure if $N \, d$ is odd.  The constraint coming from the gauge field
implies that the ``Lorentz group'' $U(d)$ representation of the field
strength of the particle is given by a rectangular Young tableaux with $q$
rows and $N$ columns.  The equations of motion \noeq{Dirac eqs} then show
that the particle is in the representation of the little group $U(d-2)$ with
$q-1$ rows and $N$ columns.

The \un{} particle can be regarded as a toy model for the $N=2$ string, to
which it is intimately related.  (The \uone{} and \utwo{} theories can be
obtained by dimensionally reducing the $N=2$ string, and then gauging the
global symmetry of the theory.)  One result that has implications for
the string is that the partition function of the particle, calculated by
performing the path integral on the circle with a flat target space, is in
perfect agreement with what one would expect from the spectrum of the
particle.  This is in contradistinction to the string, in which modular
invariance forces the string partition function to have a peculiar dependence
on the proper time $\tau$.  The fact that this does not occur in the
particle rules out several of the explanations proposed for this phenomenon.

Other unsettled issues in the $N=2$ string are whether or not it is the same
as the $N=4$ string, and whether or not it has spacetime supersymmetry and a
full $SO(D)$ Lorentz invariance \cite{allWarren}.  It is less clear here what
implications can be drawn from the particle.  One would not expect to
be able to see spacetime supersymmetry in any (non super) particle.
The general \un{} particle
defined on some \Kahler{} space certainly does not have an $SO(D)$ Lorentz
invariance.  Indeed, the spectrum will not even
fall into representations of $SO(D)$.  It is also not equivalent to a
$USp(N/2)$ particle.  However, these properties may still hold in the special
case of the $N=2$ string, for which spacetime must be hyper\kahler{} in four
real dimensions.

Some issues which we have not explored, but which should not pose any great
difficulty, are how to add masses to the \un{} theories, and how to construct
the $USp(N)$ theories, which are related to the $N=4$ strings.  Masses have
been included in the original spinning particle \cite{spart}, and in the
\on{} particle \cite{russian\on}, although only in flat space.  One should
be able to include them in the \un{} strings in the same way.  Alternatively,
one can introduce the masses by dimensionally reducing the $N=2$ string using
the Scherk--Schwarz mechanism \cite{ScherkSchwarz}.  Also, following the line
of this paper, it should not be hard to find the proposed $USp(N)$ particles,
for example by dimensionally reducing the $N=4$ string \cite{quater}.  As
with that string, these theories should be definable only on hyper\kahler{}
or quaternionic spacetimes.  A more difficult and more interesting problem is
to classify all spinning particle theories.

\newpage
{
\small
\parskip=0pt plus 2pt
\def\baselinestretch{1.}

}

\end{document}

\appendix{Extra equations}
\beqal
&\delta \, X^\mu =  -i \, \alphastar \cdot \chi^\mu \\
&\delta \, \psistaro i = D \, \alphastaro i  \\
&\delta \, \chi_\mu^{*\,i} = \nfrac1e \, G_{\mu\mubar}
                 \left( \, \dot \Xstaro \mubar
                  + i \, \psi \cdot \chistaro \mubar \, \right) \alphastaro i
                - i \, \Gamma_{\mu\rho,\sigmabar} \;
                \alphastar \cdot \chi_\rho \, \chistart i \sigmabar \comma\\
&\delta \, \chi_i^\mu = \nfrac1e \left( \, \dot X^\mu + i \, \psistaro j \,
        \chi_j^\mu \, \right) \alpha_i + i \, \Gamma_{\rho\sigma}^\mu \,
        \alphastaro j \chi_j^\sigma  \chi_i^\rho\\
&\delta  \, \chistart i\mubar =  \nfrac1e \left( \, \dot \Xstaro \mubar
        + i \, \psi_j \; \chistart j\mubar \, \right) \alphastaro i
         + i \, \Gamma_{\rhobar\sigmabar}^\mubar \,
        \alpha_j \chistart j\sigmabar \chistart i \rhobar
\eeqal

\end{document}